\documentclass[aps,showpacs,showkeys]{revtex4}
\usepackage{graphicx}
\begin{document}

\title{SN1987A and the properties of the neutrino burst}

\author{Maria Laura Costantini}
\email[]{Marialaura.Costantini@aquila.infn.it}
\affiliation{University of l'Aquila, L'Aquila, Italy and INFN}
\author{Aldo Ianni}
\email[]{Aldo.Ianni@lngs.infn.it}
\author{Francesco Vissani}
\email[]{Francesco.Vissani@lngs.infn.it}
\affiliation{INFN, Laboratori Nazionali del Gran Sasso,
Assergi (AQ), Italy}


\begin{abstract}
We reanalyze the neutrino events from SN1987A 
in IMB and Kamiokande-II (KII) detectors,
and compare them with the expectations from  
simple theoretical models of the neutrino emission.
In both detectors the angular distributions are 
peaked in the forward direction, and 
the average cosines are 2 sigma above the expected values. 
Furthermore, the average energy in KII is low 
if compared with the expectations; 
but, as we show, the assumption that a few 
(probably one) events at KII have been caused by elastic scattering 
is not in contrast with the 'standard' 
picture of the collapse and yields 
a more satisfactory distributions 
in angle and (marginally) in energy.
The observations give useful information 
on the astrophysical parameters of the
collapse: in our evaluations, the mean 
energy of electron antineutrinos is $\langle E\rangle =12-16$~MeV, 
the total energy radiated around $(2-3)\times 10^{53}$~erg, and there
is a hint for a relatively large radiation 
of non-electronic neutrino species.
These properties of the neutrino burst are not in disagreement 
with those suggested by the current theoretical
paradigm, but the data leave wide space to non-standard pictures, 
especially when neutrino oscillations are included.
\end{abstract}



\maketitle

\section{Motivations and context}\hspace{0.5cm} 

The detection of neutrinos from SN1987A marked the 
beginning of (extra)galactic neutrino astronomy 
\cite{IMBdata87,IMBdata,KIIpaper87,KIIpaper,Baksan87,MontBlanc} 
(see also~\cite{Olga,raffelt,Koshiba} for 
comprehensive reviews of SN1987A observations).
The observations of Kamiokande have 
been mentioned and recognized in 
the 2002 Nobel prize for Physics.
However, when one studies the data, one meets a number 
of surprising, unexpected or even puzzling 
features. Let us recall which are the main ones.\\
(1)~The angular distributions of the 
events seen at Kamiokande-II (KII) and at Irvine-Michigan-Brook\-haven (IMB) 
are more forward-directed than expected, for instance, the average 
cosines of the polar angles are  
$\langle \cos\theta^{\rm KII}\rangle\sim 0.3$ 
and 
$\langle \cos\theta^{\rm IMB}\rangle\sim 0.5$.\\
(2)~Also, the energy distribution of these 
two detectors seems not to be perfectly in 
agreement. In particular, 
$\langle E_{vis}^{\rm KII}\rangle$ is half  
of $\langle E_{vis}^{\rm IMB}\rangle$ (about $30$~MeV),
that is, a very marked difference even taking into account 
the different performances of the detectors.\\
(3)~Even the time distribution of the events in the two
detectors looks to be different. However, when data are combined 
the distribution in time does not contradict the current 
picture of a `delayed explosion' according to Lamb 
and Loredo analysis \cite{LambLoredo}.\\
(4)~The Mont Blanc events \cite{MontBlanc} occurred $4.5$ hours 
before the other ones. This led some Authors to consider 
two-stage scenarios for the collapse \cite{nb1}. \\
In this work, we will focus on the discussion of the 
first issue and will stress the connections with the second 
one. More in general, we believe that these 
data raise several important questions
that deserve attention, for instance: 
How likely is that the anomalies in the distributions are 
due to fluctuations, and in particular, how significant is the 
hint for some feature in the angular distributions? 
What can we learn (and what we can exclude) 
on the nature and the properties of 
the stellar collapse from these observations?

A number of recent facts, beside the general considerations
exposed above, testify the interest in 
having a fresh look at the SN1987A data:
(a)~several experimental evidences (in particular 
\cite{SuperKamiokande,SNO,Kamland}) strongly suggest 
that SN1987A neutrinos oscillated in flavor;
(b)~the expectations of the emitted neutrino radiation 
has been recently reconsidered 
\cite{Janka}, suggesting a new paradigm for the distribution 
of neutrinos and antineutrinos;
(c)~there have been improvements in the description of the 
cross section of 
\begin{equation}
\bar{\nu}_e\;  p\to e^+\; n,\ \ 
\mbox{IBD reaction (from `inverse beta decay')} 
\end{equation}
in the energy range relevant for supernova neutrinos \cite{beacom,vissani}.
Moreover, it is correct to recall that we do not understand yet 
the theory of core collapse supernovae, and therefore 
one could argue that we miss the most important 
ingredient for a proper interpretation. 
However, a reasonable working hypothesis 
is to describe the emitted neutrino radiation 
by a model with few parameters, suggested by 
the `delayed explosion' scenario proposed 
in \cite{del}, see \cite{JANKA} for a recent report.
This is the point of view we will adopt in a large 
part of the present investigation. 

We will describe and motivate in the rest of this 
Section what we assume (based on expectations and 
observations) as a reference neutrino flux.
We will discuss a standard (but updated) comparison 
of observations and expectations in Section~2, 
based on IBD hypothesis (see below), that will permit to further define 
the parameters of the model of neutrino emission.
In Section~3, we will use this model to 
analyze the angular features of the spectra, 
state the situation quantitatively, 
and consider a few alternatives 
to improve the agreement with the data.
We summarize the results obtained 
in the last Section.

\subsection{Neutrino flux\label{nf}}

A simple model of the fluxes \cite{Janka}
of supernova neutrinos 
attributes the following spectra (with three different 
average energies $E_0$) to any species 
$\nu_e$, $\nu_{\bar e}$ or $\nu_x$--$x$ 
being any among muon and tau (anti)neutrinos:
\begin{equation}
\Phi_i(E)=\frac{{\cal E}_i}{4 \pi D^2}\ 
\frac{N}{E_0^2}\ z^\alpha e^{-(\alpha+1) z}, \ \ z=E/E_0
\label{fd}
\end{equation}
where $i=e,\bar{e},x$ and 
$N={(\alpha+1)^{\alpha+1}}/{\Gamma(\alpha+1)}$.
The total fluence at the detector is 
$\int E \Phi_i(E) dE={{\cal E}_i}/4 \pi D^2$,
thus ${\cal E}_i$ is the amount of irradiated energy 
in the neutrino species~$i$ (the flux is supposed to be emitted 
isotropically) and $D$ the SN-detector distance.
Numerical calculations find that the 
time integrated flux $\Phi$, usually called `fluence',
is rather well described by this ansatz; in particular,
the deviations from a thermal shape 
are not large, and can be described as we do here 
by setting $\alpha= 3$ for all neutrino species 
($\alpha=2$ amounts to a Maxwell-Boltzmann distribution).
Finally, the meaning of $E_0$ is just the average energy of the
species considered (we will take $E_0=\langle E_{\bar e}\rangle$ 
in the following).

The total energy emitted in neutrinos 
can be estimated by simple considerations. 
In fact, the total gravitational energy irradiated is 
${\cal E}_B\sim 3 G_N M_{ns}^2/5 R_{ns}$, and 
using for the neutron star a 
mass of $M_{ns}=(1-2) M_\odot$ and a radius of 
$R_{ns}=20\mbox{ km }(M_\odot/M_{ns})^{1/3}$, we get 
${\cal E}_B\sim (1-5)\times 10^{53}$~erg. 
The amount of energy that goes in the specific flavors 
is uncertain. Since ${\cal E}_e$ is not 
very important for the observed signal (see below), 
we will always set ${\cal E}_e={\cal E}_{\bar e}$,
unless stated otherwise.
Instead, we will distinguish three cases for 
the emitted energy ${\cal E}_x$ (assumed to be equal for 
$\nu_\mu$, $\nu_{\bar \mu}$, $\nu_\tau$ and $\nu_{\bar \tau}$,
so that ${\cal E}_B={\cal E}_e+{\cal E}_{\bar e}+4{\cal E}_x$)
that, as we will see, plays a more important role:
\begin{enumerate}
\item ${\cal E}_x={\cal E}_{\bar e}$: This is the 
so-called `equipartition', often adopted in theoretical analyses.
\item ${\cal E}_x={\cal E}_{\bar e}/2$: This is the  
case when a large part of the radiation goes in electron neutrinos.
\item  ${\cal E}_x=2 {\cal E}_{\bar e}$: Finally, in this  
case most of the radiation goes in muon or tau neutrinos.
\end{enumerate}

The average energies are important parameters. 
$\langle E_x\rangle$ is greater than $\langle E_{\bar e}\rangle $,
but the amount of hierarchy found in modern calculations is not
very large. A typical ratio is in the range $1-1.2$.
In the following, we will assume (unless stated otherwise)
$\langle E_x\rangle=1.1\langle E_{\bar e}\rangle $.
The average energy of the electron neutrinos $\langle E_{e}\rangle $ instead 
is not of crucial importance for the observed signal. 
It can be evaluated by prescribing a condition on the 
emitted lepton number $\Delta L_e=N(\nu_e)-N(\bar{\nu}_e)$ where
$N(\nu_e)={\cal E}_e / \langle E_e\rangle $ and similarly for the 
antineutrinos; we will assume that the electrons contained in  
one solar mass of iron are converted in neutrinos.

The crucial parameters needed to 
describe the neutrino signal are the antineutrino 
average energy, 
\begin{equation}
E_0\equiv \langle E_{\bar e}\rangle
=12-18\ {\rm MeV}\; \; \ \ \  \mbox{(expected)} 
\label{ex_e0}\label{3}
\end{equation}
and the total energy irradiated in antineutrinos 
\begin{equation}
{\cal E}_{\bar e}=(2-10)\times 10^{52}\ {\rm erg}\ \ \ \ \ \ \ 
\mbox{(expected)} 
\label{ex_e}\label{4}
\end{equation}
Both of them have considerably uncertainties, especially the second one.
For this reason, the uncertainty in the distance of the 
supernova is usually considered unimportant; here, we will 
assume $D=52$~kpc, and discuss this point later.

\subsection{Impact of neutrino oscillations\label{sec:sec}}

Motivated by the solar and atmospheric results, we assume 
that the three neutrinos $\nu_e$, $\nu_\mu$ and $\nu_\tau$
have mass and mix among them. 
Following simple minded theoretical expectations,
we will further assume in most of this paper  
that the heaviest state is separated by 
$\Delta m^2_{31}\approx 2.5\times 10^{-3}$~eV$^2$ from the other two,
whose splitting is $\Delta m^2_{21}\approx 7\times 10^{-5}$~eV$^2$.
The known mixing angles are $\theta_{23}\approx 45^\circ$, 
$\theta_{12}\approx 34^\circ$, while $\theta_{13}$ is unknown 
but presumably it is not 
very small (we take $\approx 6^\circ$ when needed, but 
its impact on the oscillations is usually of minor importance). 
With these parameters, the emitted 
fluxes from SN1987A, described in Sect.\ref{nf},
should be modified to 
account for the MSW effect in the 
star~\cite{raffelt_neubig,dutta,fogli,burrows_osc,smirnov}
(among first papers 
on the topic, we recall 
\cite{Mikheev_smirnov,Kuo,Minakata,Arafune,Notzold}):
\begin{equation}
\left\{
\begin{array}{l}
\Phi_{e}\to P_{ee} \Phi_e + (1-P_{ee}) \Phi_x \\
\Phi_{\bar{e}}\to P_{\bar{e}\bar{e}} \Phi_{\bar{e}} + (1-P_{\bar{e}\bar{e}}) 
\Phi_x 
\end{array}
\right.
\label{osc}
\end{equation}
where the two probabilities of survival are 
$P_{\bar{e}\bar{e}}=\cos^2\theta_{12}\approx 0.7$ and
$P_{ee}=\sin^2\theta_{13}\approx 0$. 
(We will discuss other possibilities, as a very small 
$\theta_{13}$, or an `inverted' mass hierarchy, 
in Sect.\ref{sec:where}.)
The MSW effect of the 
Earth modifies $P_{\bar{e}\bar{e}}$ by a minor amount \cite{nb2}.

Two remarks are in order: (1)~It is difficult to conceive that
oscillations did not occur; for instance, the MSW effect related to 
solar $\Delta m^2$ happened unless there was a drastic modification of the 
mantle of the star for densities around 10~gr/cc, which seems unlikely. 
(2)~The effects for $\bar{\nu}_e$ are of about 30~\%, while those
for $\nu_e$ can be much larger; for instance, in the framework 
described above, the observed $\nu_e$ flux corresponds almost exclusively
to emitted $\nu_\mu$ or $\nu_\tau$.  This is the reason why ${\cal E}_x$
has an important role for the signal seen in terrestrial detectors.

\section{The IBD hypothesis}

\begin{figure}[t]
\includegraphics[width=.92\textwidth,angle=270]{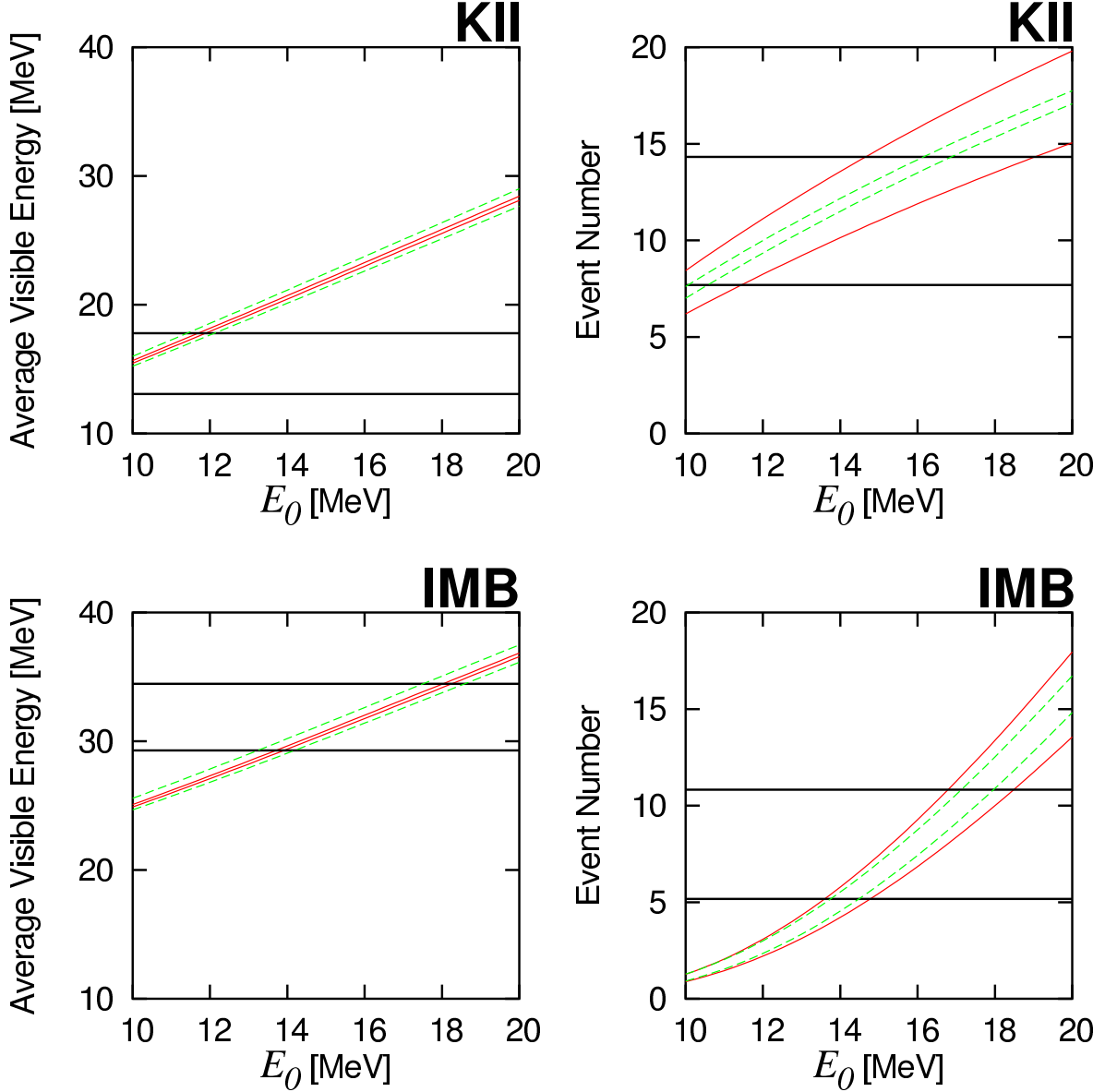}
\caption{\em Comparison of observations 
(horizontal strips) and expectations calculated in the IBD hypothesis.
The left panels show the average visible energy, the right ones 
the number of events. The upper panels are for KII, 
the lower ones for IMB. For any panel, we show 4 expectations curves.
The continuos (red) ones 
correspond to a variation of the theoretical parameter 
${\cal E}_x$ (=energy radiated in  $\bar{\nu}_{\mu,\tau}$)
in the range $(2-6)\times 10^{52}$~erg. The  dashed ones (green) 
correspond to a variation of the theoretical parameter 
$\langle E_x\rangle$ 
(=average energy of the emitted $\nu_\mu$,  $\nu_{\bar \mu}$, 
$\nu_\tau$ and $\nu_{\bar \tau}$) in the range 
$(1-1.2)\times E_0$. 
}\label{fig1}
\end{figure}

In the model previously described and with the expected values of the 
parameters, it is a fact that most of the events expected at 
KII and IMB are due to the inverse beta decay process. This is the 
reason why several analyses adopted the simplifying hypothesis 
that {\em all} events come from IBD (see e.g.\ 
\cite{janka_hille}). We begin by repeating 
such a more-or-less standard analysis, with three specific aims: 
(1)~stressing observables with a clear physical meaning (rather than 
attempting a global analysis of the data);
(2)~discussing how the  various data fit in the theoretical picture;
(3)~getting more specific values of the parameters of neutrino emission.
The calculations of the expectations are quite simple. 
In a detector with $N_p$ protons and with
detection efficiency $\epsilon(E)$ (function of the positron energy) 
one integrates the differential event rate 
\begin{equation}
\frac{d N_{ibd}}{dE_\nu dE} = N_{p}\; \epsilon(E)\; \Phi_{\bar e}(E_\nu)\; 
\frac{d \sigma_{ibd}}{dE}(E_\nu,E)
\end{equation}
over the allowed range, obtaining the value of the observable 
of interest--e.g., the visible energy in \v{C}erenkov detectors $E_{vis}=E$
(note that the fluence $\Phi$ should be thought as differential
in the transversal surface {\em and} in $E_\nu$). 
In Figure~\ref{fig1}, we show the value of two simple 
but interesting observables: the mean energy and the number of events. 
In this Figure, the effect of varying $\langle E_{\bar e}\rangle$,
$\langle E_{x}\rangle $ and ${\cal E}_x$ in  
certain ranges are illustrated. For IMB, we 
reduced the expected number of events by 13\% to
take into account the dead-time occurred during 
the detection of the burst~\cite{IMBdata}. Let us comment the
results in some detail:
\paragraph{Average visible energy}
This observable has the advantage of being 
independent of the total energy emitted, and of 
having relatively small errors:
\begin{equation}
\delta E_{vis}=\sqrt{\frac{\langle E^2_{vis}\rangle-\langle E_{vis}\rangle^2}{N}}=
\left\{ 
\begin{array}{l} 
2.4\ \mbox{MeV at KII}\\
2.6\ \mbox{MeV at IMB}
\end{array}
\right.
\end{equation}
in the IBD hypothesis 
$E_{vis}$ is the energy released by the positron.
While IMB points to a range of values nicely consistent with expectations,
compare with eq.(\ref{ex_e0}), the data of KII point to somewhat lower values.
Note that we discard from the analysis the sixth event of KII, since it
has $N_{hit}=16$, below the threshold $N_{hit}=20$ 
of software analysis~\cite{KIIpaper}. 
Indeed, it should be remarked that the lower energy 
events are those for which pollution from background is 
more likely; in particular,
in the window of 12 sec in which the supernova neutrinos 
have been detected at KII, we estimate an average of 
about 2 background events~\cite{KIIpaper}. 
From Figure~\ref{fig1}, one sees that the impact of a variation of 
$\langle E_{x}\rangle$ and ${\cal E}_x$ on the 
expectation for the average visible energy is not large.
\paragraph{Number of events}
The observables $N^{\rm KII}=11$ and $N^{\rm IMB}=8$
have large Poisson errors,  
but permit to estimate the energy emitted from the supernova (whereas,
the previous observable is not useful for this purpose). In the 
two plots on the right of Figure~\ref{fig1}, the energy emitted 
in any species of neutrino is chosen by default to be  
$4\times 10^{52}$~erg.  We see that the agreement 
with the expectations is not bad, and also that the impact of a 
variation of ${\cal E}_x$ is not fully negligible. This is easy to understand
and to keep into account; indeed, the signal scales roughly as 
$0.7 {\cal E}_{\bar e}+0.3 {\cal E}_{x}$, thus a variation in ${\cal E}_x$ 
can be well simulated by a variation of the total emitted 
energy. 

\paragraph{Summary}
Using these results as a guide, we further specify 
the parameters of the model and assume
\begin{equation}
E_0\equiv \langle E_{\bar e}\rangle=14~\mbox{MeV},\ \ \ 
{\cal E}_{\bar e}=4\times 10^{52}\ \mbox{erg} 
\label{choice}
\end{equation}
These values should be thought as compromises 
between contrasting needs. 
For instance, KII energy spectra suggest 
lower values of $E_0$,  whereas in order
to reproduce the number of events at central values, we need 
$E_0\sim 18$~MeV and ${\cal E}_{\bar e}=3\times 10^{52}$~erg.
In view of this situation, we find it reassuring that the values shown 
in eq.(\ref{choice}) are in accordance with the expectations from a 
`standard' collapse (as defined in eqs.~(\ref{3}) and (\ref{4})). 
 
We conclude this first part of the analysis reassessing  
that within the `standard' model of the collapse  
and with the parameters of eq.~(\ref{choice}),  the 
observed average visible energy at  KII looks a bit 
small (see again Figure~\ref{fig1}).

\section{What is the meaning of the forward events?}
\begin{table}[t]\setlength{\tabcolsep}{1.5pc}
\caption{\em Approximate coefficients 
of the angular distributions for IBD using eq.~(\ref{choice}).
 \label{tab1}}
\begin{center}\begin{tabular}{|l|c|c|c|}
\hline
               & $a_0$ & $a_1$ & $a_2$ \\ \hline
KII            & 0.499 & 0.030 & 0.003 \\ 
IMB            &  0.495 & 0.104 & 0.014 \\ 
IMB with bias  &  0.492 & 0.154 & 0.024\\  \hline
\end{tabular}\end{center}\end{table}

In this Section we study the angular distribution 
of the events from SN1987A.
Thus, we select the events from the two water 
\v{C}erenkov detectors operative at that time, and 
use the data from~\cite{IMBdata} for 
IMB, and those from~\cite{KIIpaper} for KII. 
Both angular distributions are 
rather forward-directed. To state this more precisely, we calculate 
the average angles:   
$\langle \cos \theta^{\rm KII} \rangle =
0.29 \pm 0.27$ and $\langle \cos \theta^{\rm IMB} \rangle =
0.48 \pm 0.34$. Here we have used a weighted average and the corresponding
standard deviation errors.

\paragraph{Beyond the IBD hypothesis}  We can compare the 
data with  the expectations from the IBD hypothesis. Using 
parameterized angular distributions 
\begin{equation}
dN/dz = a_0 + a_1 z + a_2 z^2, \ \ \ \mbox{where }z=\cos\theta\mbox{ and }
a_i\mbox{ as in Tab.\ref{tab1}}
\label{eq3}
\end{equation}
obtained from \cite{vissani} 
we find that both central values are above 
the expected ones: 2.3$\sigma$ for IMB and 1.7$\sigma$ for KII. 
This conclusion is in agreement with~\cite{beacom}.
In this study, we adopt the model defined in the previous Section
with the parameters of eq.~(\ref{choice}).
We checked that a variation of these parameters is not 
crucial for the conclusions, while ${\cal E}_x$ is of greater 
importance. It is simple to explain the reason: 
The only type of events that is strongly forward (and thus is able 
to affect the angular shape of the distribution)
are those from
\begin{equation}
\nu_i\; e\to \nu_i\; e,\ \ \ 
\mbox{ES reaction (from `elastic scattering'),}\ \ \ \ 
i=e,\bar{e},x 
\label{es}
\end{equation}
This reaction receives contributions from all neutrino types, 
and $\nu_e$ gives the largest one. But due to oscillations, 
eq.~(\ref{osc}), the observed $\nu_e$ flux is originally due to $\nu_x$; 
this implies the relevance of ${\cal E}_x$, 
namely, the energy emitted in $\nu_{\mu,\tau}$.
The hypothesis that one or more forward peaked elastic 
scattering events could be present 
in the data samples of IMB and KII has been 
already considered in the past, see  
e.g.~\cite{spergel,krauss,LoSecco,Olga,MIK,
malguin}. 
In this analysis, however, we update 
the angular distributions for IBD events 
and the model for neutrino emission, 
compare different statistical inferences
and include oscillations with recently measured parameters.

\begin{figure}[t]
  \begin{center}
  \mbox{\includegraphics[width=0.56\textwidth]{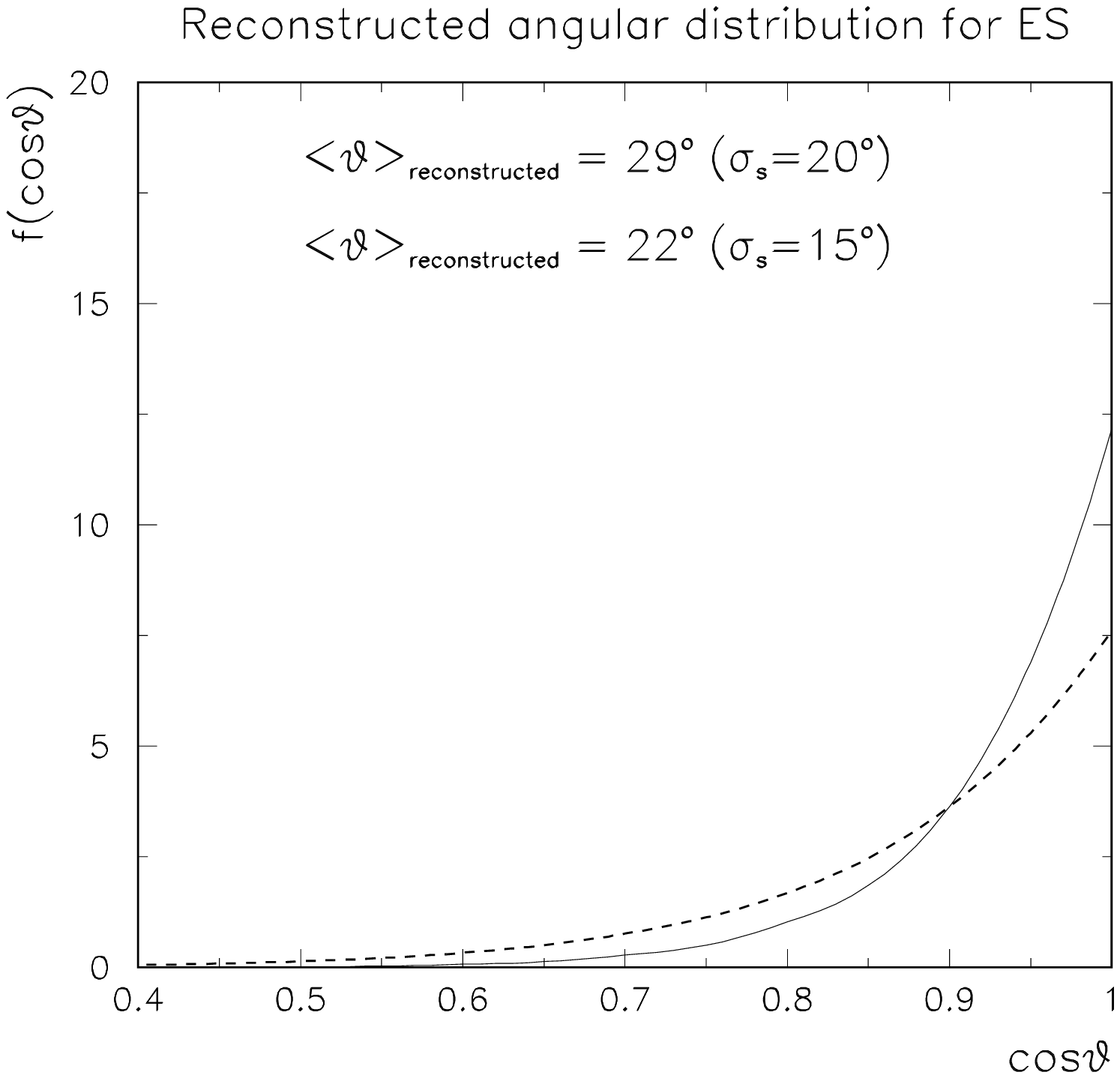}}
  \end{center}
\vskip-1.4cm
  \caption{\em{Reconstructed angular distribution for elastic scattering 
events in KII.}}\label{fig:smearing}
  \end{figure}

\paragraph{Instrumental effects}
A point to take into account is that the angular distributions
(and in particular the one of ES)
are modified in an important manner by instrumental effects.
This is due to multiple scattering and limited angular resolution of the 
detectors, and it is called  `smearing' of the angular distributions.
In order to account for this, we use the following 
distribution~\cite{Danuta}:
\begin{equation}
\rho_{sm}(\cos\theta_s) d\cos\theta_s = 
N \, e^{-\theta^2_s / 2\sigma_s^2}\, \sin\theta_s   d\theta_s\, \label{eq5}
\end{equation}
where $\theta_s$ is the angle of smearing 
and $\sigma_s$ is a measure of  the effect;  
$N$ in eq.~(\ref{eq5})  is just 
a normalization factor. 
For KII, where the smearing is slightly more important, 
we choose  $\sigma_s$ in such a way 
that the mean angle $\overline{\theta}$ from~(\ref{eq5})
corresponds to the mean error ${\delta\theta}$ determined from 
the data~\cite{KIIpaper}. For the whole set of data we have
$\delta{\theta}\sim 25^\circ$, while considering 
only data with $\theta \leq 30^\circ$ we find
$\delta\theta \sim 18^\circ$.
As a consequence we decide to study the two cases, 
when $\sigma_s=15^\circ$ and $\sigma_s=20^\circ$, for which 
$\overline{\theta} \sim 18^\circ$ and $\sim 25^\circ$, 
respectively.
Based on similar considerations~\cite{IMBdata}
we set $\sigma_s=16^\circ$ in IMB.
Using eq.~(\ref{eq5}) and $\rho_{es}$ e.g.\ from~\cite{bahcall}
we can determine the reconstructed angular distribution as follows:
\begin{equation}
\rho_{es}^{rec}({\bf n} \cdot {\bf m }) = \int d^2{\bf p}\ 
\rho_{sm}({\bf m } \cdot {\bf p}) \
\rho_{es}({\bf n } \cdot {\bf p }) 
\label{eq6}
\end{equation}
where ${\bf n}$, ${\bf{m}}$ and ${\bf p }$ are unitary vectors 
for the SN, the reconstructed and the emitted direction respectively,
${\bf n } \cdot {\bf m}=\cos\theta$, 
${\bf m } \cdot {\bf p}=\cos\theta_s$, 
${\bf n } \cdot {\bf p}=\cos\theta\cos\theta_s+
\sin\theta\sin\theta_s\cos\phi_s$, 
and $d^2{\bf p}=d\cos\theta_s d\phi_s /4\pi$ 
is an element of solid angle from which the
signal receives a contribution. The reconstructed distributions
we obtain for KII are plotted in Figure~\ref{fig:smearing}.  
Since the angular distribution $\rho_{es}$ of ES 
is rather narrow (especially when taking into account detector
efficiencies) the reconstructed angular distribution of ES 
is mostly dictated by instrumental effects.

\subsection{Angular distribution of IMB}
The normalized positron angular distribution for inverse beta decay 
is usually taken in the simple approximation 
$dN/d\cos\theta = 0.5 + a \cos\theta$;  
in particular, in the IMB report~\cite{IMBdata}
it is assumed $a \sim 0.07$. 
In the same paper it is pointed out that to account for
the experimental polar-angle efficiency, 
one can introduce a 10 \% angular bias.
This is equivalent to replace 
$dN/d\cos\theta \rightarrow (1+0.1\cos\theta)\ dN/d\cos\theta$.
We use the improved cross-section for 
IBD from~\cite{vissani} to determine the parameters  
$a_i$ ($i=0,1,2$) in Tab.~\ref{tab1}, that enter in the 
angular distribution of eq.(\ref{eq3}). 
We notice that $a_i$'s in Tab.\ref{tab1} do not depend
significantly on the assumed mean energy $E_0$.

\begin{figure}[t]
  \mbox{\includegraphics[width=0.56\textwidth]{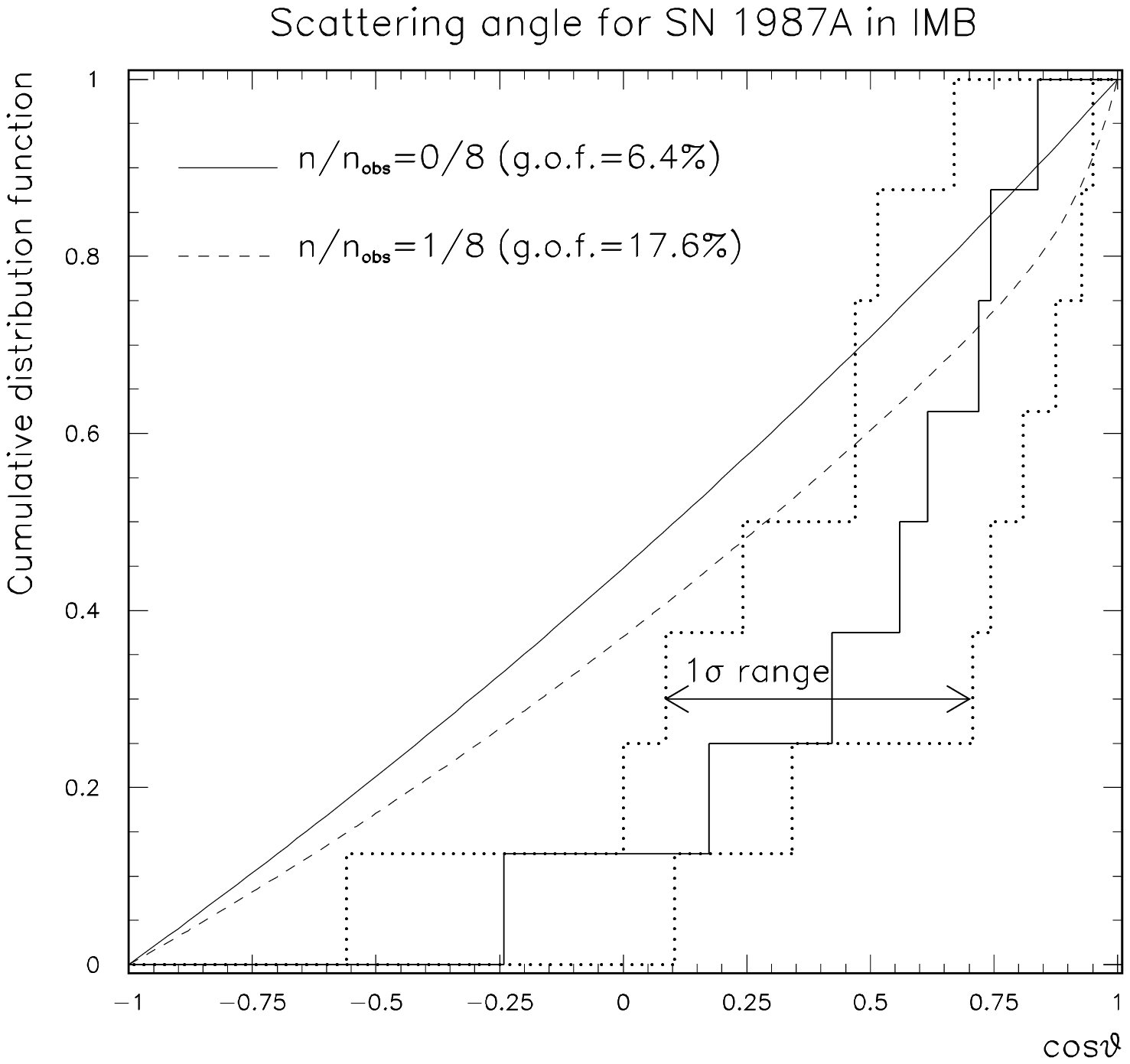}\hskip-12mm
  \includegraphics[width=0.56\textwidth]{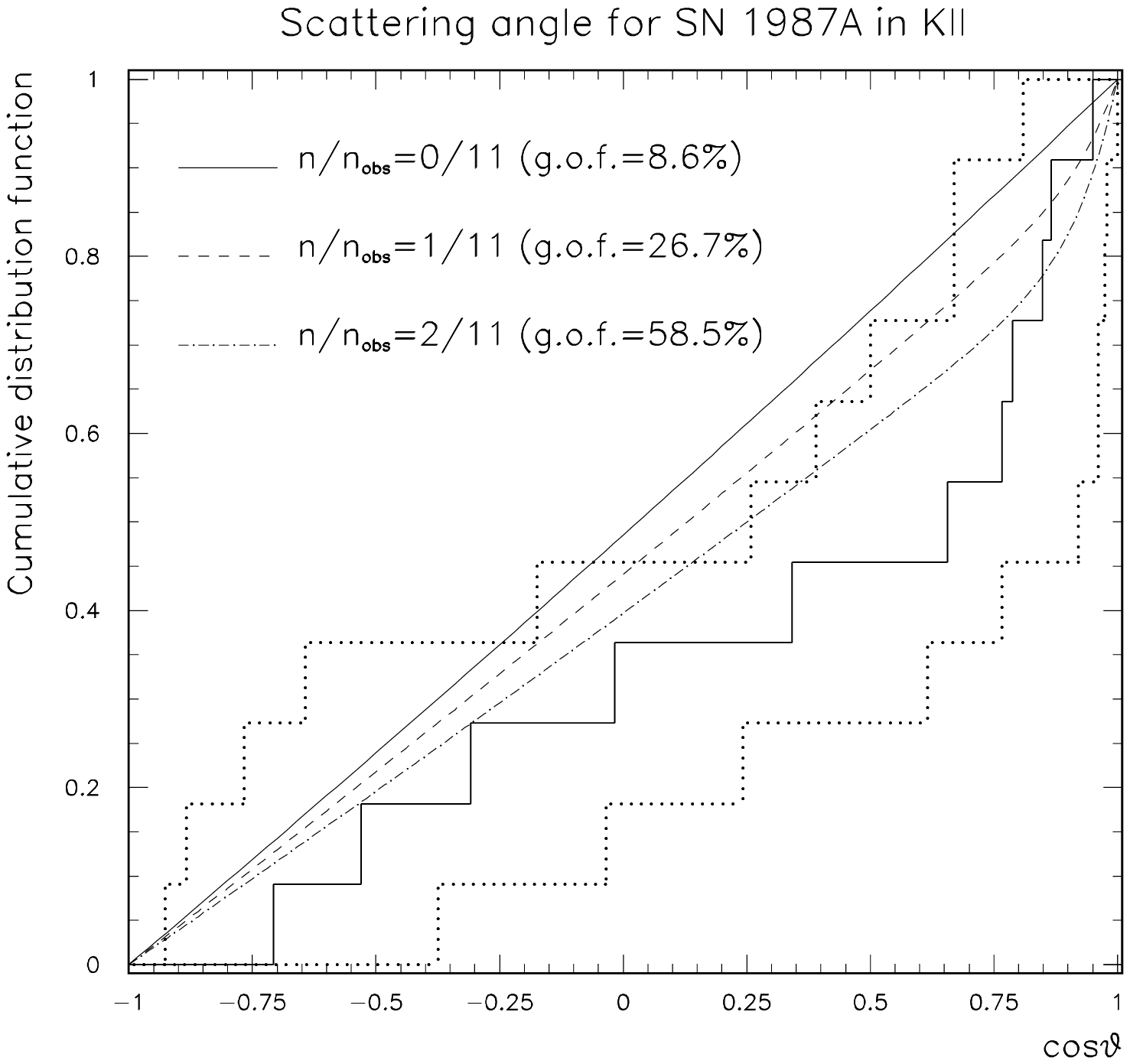}}
\vskip-1.4cm
  \caption{\em{Theoretical and experimental 
angular distributions in 
IMB (left panel) and KII (right panel). The 1~sigma
range and the goodness of fit figures are also shown. 
For KII, we adopt $\sigma_s=15^\circ$.}}\label{fig:distrangolare}
  \end{figure}

We have used eq.~(\ref{eq3}) to test
the hypothesis the data from IMB come from IBD events, 
employing the Smirnov-Cramer-Von Mises (SCVM) 
statistics~\cite{SCMtest}.  
As shown in the left panel 
of Figure~\ref{fig:distrangolare} the 
goodness of fit (g.o.f.) for this 
hypothesis is equal to 6.4\%.
The improved IBD angular distribution changes 
the previous result (4.5\%~\cite{IMBdata})
by only a small amount due to the poor statistics. 
However, the importance of using the improved angular distribution 
is evident when we compare the old significance 
without angular bias with the new one, since 1.5~\%~\cite{IMBdata}  
increases to 4.2~\%. In the same Figure we show the cumulative distribution in
the hypothesis of having 1 ES event in IMB.

To study the possibility to have a small contribution of 
Elastic Scattering (ES) events in IMB 
and later in KII we have exploited the Maximum
Likelihood (ML) method~\cite{bookEML}. In this framework the 
likelihood function is written:
\begin{equation}
 L\left(\frac{n}{n_{obs}}\right) =  \prod_{i=1}^{n_{obs}}
 \rho\left( \cos\theta_i; \frac{n}{n_{obs}} \right) \, \label{eq4}
\end{equation}
where $n/n_{obs}$ is the parameter which measures 
the fraction of ES events, $n_{obs}$ being the
total number of experimentally observed events for 
the SN. The angular distribution $\rho(\cos\theta_i;n/n_{obs})$
can be written as 
$\rho(\cos\theta;n/n_{obs})=(n/n_{obs}) \rho_{es}^{rec}(\cos\theta) + 
(1-n/n_{obs}) \rho_{ibd}(\cos\theta)$, where  
$\rho_{es}^{rec}$  and $\rho_{ibd}$ are 
the angular distributions for ES and IBD, respectively. 
It turns out that in IMB the best-fit is
found for $n/n_{obs} = 0$. The effect of the smearing is not 
particularly important in IMB.
In order to determine an upper limit on the likelihood parameter we have
built a posterior probability distribution function
(p.d.f.) by normalizing the likelihood function 
and considering a uniform prior p.d.f.\ 
which is equal to one for $n/n_{obs} \ge 0$, zero elsewhere. 
It turns out that $n/n_{obs} < 0.12$ at 
68.3\% C.L., namely the IMB angular distribution admits 
one ES event at most.

\begin{figure}[t]
  \begin{center}
  \mbox{\includegraphics[width=0.6\textwidth]{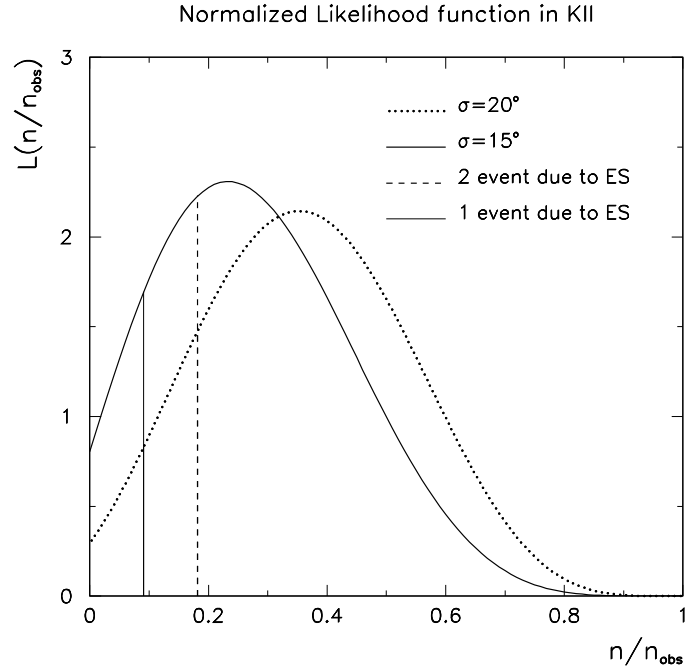}}
  \end{center}
\vskip-1.4cm
  \caption{\em{Normalized likelihood function versus 
the fraction of expected ES events in KII. 
  For the smearing of the angular distribution we show the cases 
$\sigma=20^\circ$ and
  $\sigma=15^\circ$.}}\label{fig:likelihood}
  \end{figure}

\subsection{Angular distribution of KII}

As stated above we consider $n_{obs}$=11 out of 12 
candidate events~\cite{KIIpaper} and assume the
event number 6 due to background. 
In Figure~\ref{fig:smearing} we show 
the reconstructed angular distributions for the cases $\sigma_s=15^\circ$ 
and $\sigma_s=20^\circ$.
The smearing effect in KII plays an important role
(without smearing, $\langle \theta \rangle = 10^\circ$). 
Using the ML from eq.~(\ref{eq4}) we have computed the 
likelihood ratio $L(n/n_{obs})/L_{max}$ with $n=0,1,2,3$ to
quantify the probability to have zero, one or 
more ES events on the basis of the angular distribution. 
In Figure~\ref{fig:likelihood} we show the normalized ML 
function against $n/n_{obs}$. Minimizing $-\ln L(n/n_{obs})$ 
the best-fit is found for $n/n_{obs} = 0.35 \pm 0.20$(1$\sigma$) 
for $\sigma_s=20^\circ$ and $n/n_{obs} = 0.23^{+0.21}_{-0.18}$(1$\sigma$) 
for $\sigma_s=15^\circ$ . So, the ML test
suggests that a few  ES events are present in the KII
dataset.

As for IMB we have exploited
the SCVM test (see Figure~\ref{fig:distrangolare}). 
Moreover, 
we have worked out the probability to have $n=0,1,2,3$ ES scattering 
events using the expectations based on the SN model described above. 
For this latter case we have written the probability to have $n$ ES 
events out of a total of $n_{obs}$=11 as the product 
of two Poisson distributions, that is equal to:
\begin{equation}
P(n) = \frac{e^{-n_{exp}}}{n_{obs}!}\ 
n_{exp}^{\ n_{obs}}\times B_p(n,n_{obs}) \, \label{eq8}
\end{equation}
where the first factor is a Poisson 
distribution with a mean value $n_{exp}=n_{es} + n_{ibd}$
and the second one a Binomial distribution with a
trial probability $p=n_{es}/n_{exp} \sim 0.03$; for instance, for the equipartition scenario we found 11.9 IBD
events and 0.39 ES events.
(Incidentally, one should 
notice that the calculation of the ES number of events is very sensitive to the
experimental efficiency at the lowest measurable energies).
In Tab.~\ref{tab2} we summarize our results. Similar calculations 
were made for IMB.

\begin{table}[b]\setlength{\tabcolsep}{1.5pc}
\caption{\em Expected probabilities that zero, one or more events 
in KII are due to ES, for 
$\sigma_s=15^\circ$ (upper part) 
and $\sigma_s=20^\circ$(lower part). The first line shows the 
{\em a priori} expectation from the model of eq.(\ref{choice}), 
while 
the second and the third line use the information 
from the observed angular distribution.
 \label{tab2}}
\begin{center}\begin{tabular}{|l|c|c|c|c|}
\hline
& $n=0$ & $n=1$ & $n=2$ & $n=3$\\ \hline\hline
from SN model & 50.5\% & 38.9\% & 9.7\% & 0.9\%\\ 
g.o.f.\ from SCVM  & 8.6\%  & 26.7\%  & 58.5\%  & 81.4\%\\
likelihood ratio $L(n)/L_{max}$   & 0.35 & 0.73 & 0.97  & 0.98  \\  
\hline\hline
from SN model & 52.3\% & 37.5\% & 9.2\% & 1.0\%\\ 
g.o.f.\ from SCVM  & 8.6\%  & 24.9\%  & 53.8\%  & 87.6\%\\
likelihood ratio $L(n)/L_{max}$   & 0.14 & 0.39 & 0.69  & 0.92  \\ \hline
\end{tabular}\end{center}
\end{table}

\subsection{Summary for the `standard' scenario and remarks}
It is instructive to compare here the 
outcomes of the two statistical tests
we used: the Smirnov-Cramer-Von Mises (SCVM) 
and the maximum likelihood (ML).
The comparison with the data follows completely different strategies:
the ML method is `local' in the sense that it 
profits of events that fall under the ES bell 
(of Fig.\ref{fig:smearing}), 
while the SCVM is `global' in the sense that it 
tries to minimize the maximal distance between
the theoretical curve and the observed one.
For IMB the SVCM test suggests more elastic 
scattering events than the ML test 
does (the reason can be understood from the right-most part 
of figure~\ref{fig:distrangolare}a:  
the most forward event has polar angle $33^\circ\pm 15^\circ$,
thus its central value is  not forward enough 
to suggest an ES event).
Instead, for KII the two tests give very similar indications.

Next, we take into account also 
the theoretical expectation on the number of ES events,
and use it together with the results from the angular 
distribution.
We combined the information from the SN model 
and that from the SCVM analysis in
Tab.~\ref{tab2}, multiplying the probabilities and normalizing
the resulting distribution to one. The results for KII 
are shown in Tab.~\ref{tab3}, 
and the combined probabilities are given 
for the case $\sigma=20^\circ$ (the case
$\sigma=15^\circ$ gives about the same result).
In particular, from Tab.~\ref{tab3} we see that one ES in KII can be 
accepted at about the same level we could accept
zero events. Moreover, even the probability to have two ES 
events is indeed not negligible. Repeating the exercise 
for IMB, we find that at `equipartition' (${\cal E}_{\bar e}={\cal E}_x$)
the combined probability to have zero (one) events 
is 80~\% (19.9~\%).

\begin{figure}[b]
  \begin{center}
  \mbox{\includegraphics[width=0.6\textwidth]{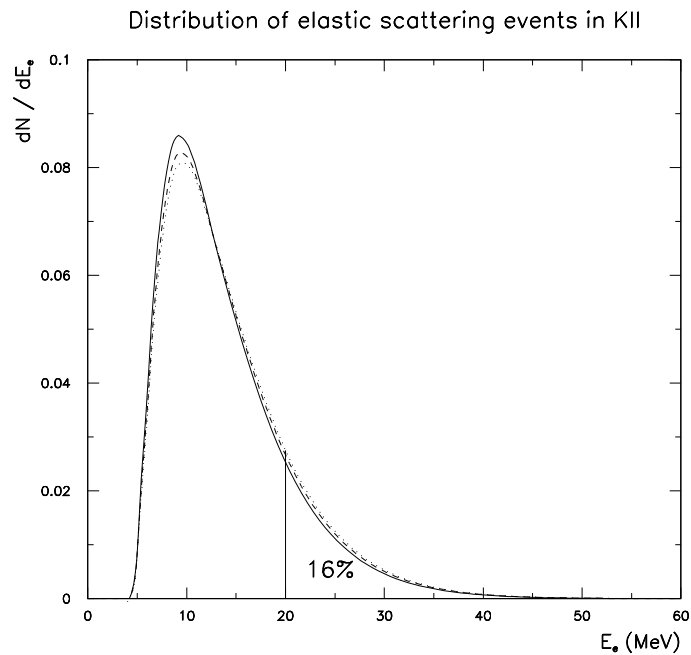}}
  \end{center}
\vskip-1.4cm
  \caption{\em{Expected energy distribution of scattering 
events in KII, for 
${\cal E}_x=(0.5,1,2) \times{\cal E}_{\bar e}$. Also shown is the 
probability to have one event with energy larger 
than 20~MeV.}}\label{fig:distr}
  \end{figure}

Some remarks are in order.\\
(1)~The data of KII show that
the most directional events have energies above 20 MeV. 
Taking this experimental fact into account, we checked
 the probability to have events with $E_e \ge 20$~MeV for the 
three scenarios of SN
considered. As shown in Figure~\ref{fig:distr} this 
probability is about 16\%. So, it seems not unlikely \cite{nb5} 
 from this point of view to have measured ES events with energies 
above 20~MeV.\\
(2)~The presence of one or more ES events in KII dataset 
goes in the right direction to explain the disagreement between
IMB and KII average energies. However the effect is admittedly small,
since for instance 1 ES event that produces $20$~MeV 
of observable energy originates from a neutrino with 
larger energy, but just of $5$~MeV on average. If the number of ES
events in KII is larger, this could become more important.

\subsection{Speculations\label{sec:where}}
\begin{table}[t]\setlength{\tabcolsep}{1.5pc}
\caption{\em  Relative percentage probabilities to have 
a given number of ES events in KII dataset, estimated from
observed angular distribution and theoretical 
expectation on the fluxes, for 3 hypotheses on ${\cal E}_x$. 
 \label{tab3}}
\begin{center}\begin{tabular}{|l|l|l|l|l|}
\hline
 & $n=0$ & $n=1$ & $n=2$ & $n=3$ \\ \hline\hline
${\cal E}_x={\cal E}_{\bar e}/2$ & 52.3\%&   37.5\%  &   9.2\%  &   1.0\% \\ 
${\cal E}_x={\cal E}_{\bar e}$ & 40.0\%  &  42.3\%   &  15.3\%  &   2.5\% \\ 
${\cal E}_x=2\; {\cal E}_{\bar e}$ & 29.0\%  &  43.6\% &   22.3\%  &   5.1\% \\ \hline
\end{tabular}\end{center}
\end{table}

It is interesting to consider at this point some speculative 
scenarios, to investigate the question under which conditions 
we can increase the expected number of ES events:
\begin{itemize}

\item A distinguished astrophysical 
possibility is that there are main departures 
from a `standard' collapse, and a large part 
of the emitted energy is not seen by inverse beta decay. 
Let us assume as an extreme case that the electron neutrinos
have an average energy of $40$~MeV and carry an energy of 
$1.5\times 10^{53}$~erg \cite{nb6}. 
The calculation reveals that the 
increase is not much larger: In KII, we expect $N_{dir}=0.59$
rather than $N_{dir}=0.39$. The reason is simply that 
oscillations transform the $\nu_e$ into $\nu_\mu$ and 
$\nu_\tau$, and the interaction cross section of 
these neutrinos is smaller.

\item Another possibility is to study which adjustment of
the `standard' scenario goes in the right direction. 
In particular, one can suggest 
that the $\nu_x$ are more energetic than what we assumed.
This does not help for the number of events, but helps a bit 
to explain the fact that the directional events are among 
the most energetic ones. 

\item The uncertainties in 
oscillations provide us with another degree of freedom. 
It seems to us very difficult to 
avoid the occurrence of MSW oscillations completely,
but if $\theta_{13}$ is very small, we could get 
$P_{ee}=0.3$.
The result $P_{ee}=\sin^2\theta_{12}=0.3$
assumes that the mantle of the star at densities of about 
10 gr/cc was not essentially modified
by the pre-collapse events. The opposite case seems unlikely,
but one could get $P_{ee}=1-\sin^22\theta_{12}/2\sim 0.6$. 
However, this does not help to increase $N_{dir}$ with the `standard'
scenario, and it is of limited use to invoke 
non-standard scenarios with energetic $\nu_e$'s, since in this case 
the reaction with oxygen are also called into play, 
see \cite{Haxton} and \cite{Burrows_gandhi}. Another
case arises if  $\theta_{13}$ is `large' when the neutrino 
mass spectrum is inverted, rather than `normal' as considered
previously in the text.
In fact, in this hypothesis
the IBD events are due to the flux $\Phi_x$, and 
the flux of $\nu_e$ (important for ES events) is $0.3 \Phi_e +0.7\Phi_x$.
Thus, we are interested to consider 
the possibility of a large ${\cal E}_e$ to increase the number of ES events, 
or the possibility that  
${\cal E}_{e}>{\cal E}_x$, since in this way we reduce  
the number of IBD events more than the ES events.

\item A more drastic attitude 
is to abandon completely the `standard' idea of the collapse. 
A  specific suggestion made in \cite{malguin}
is that a large amount of neutrino radiation comes from 
$\pi^+\to \mu^+ \nu_\mu$ decay. 
The largest contribution to scattering events comes from the 
electron neutrinos, that, due 
to oscillations   originate exactly from $\nu_\mu$
(in fact, due to the loop-induced 
difference of potential
between $\tau$ and $\mu$ neutrinos $\nu_\mu \to \nu_1$ 
and a $\nu_e$  happens to be produced with probability 
$P_{\mu\to e}=0.7$).
The $\nu_\mu$ are monochromatic with energy $29.8$~MeV.  
If the energy injected in $\nu_\mu$ is $5\times 10^{53}$~erg
we get about 1 ES event in IMB and 3 ES events in KII
(with a few additional Oxygen events). Apart from the obvious 
objection that we need to produce $10^{58}$ (!) pions, we 
are left with the problem to explain the main part of the signal
(that in the standard interpretation is attributed to IBD).
\end{itemize}
In summary, we see that there are several interesting possibilities 
and the fact that we do not have a 
definitive theory of the collapse motivates their consideration,
even though our cursory investigation seems to suggest 
that it is not so easy to produce radical modifications
of the `standard' paradigm.

\section{Summary and discussion}

We reanalyzed the neutrino signal of SN1987A 
in IMB and KII detectors in the light of new facts. 
In particular, occurrence of three neutrino 
oscillations (as defined in Sect.\ref{sec:sec}) implies
that the observed $\bar{\nu}_e$ have a 
30 \% contribution from the original 
$\bar{\nu}_{\mu}$ or $\bar{\nu}_\tau$, while 
the observed electron neutrinos ${\nu}_e$ are
practically purely 
${\nu}_{\mu}$ or ${\nu}_\tau$ at the production.
Thus, the `standard' picture of the neutrino fluxes
implies that the inverse beta decay signal 
is mostly sensitive to originally produced $\bar{\nu}_e$, 
whereas an important contribution to 
the directional signal comes from ${\nu}_{\mu,\tau}$.

The hypothesis that most of the events were due to 
inverse beta decay is in agreement with the data, 
although the presence of one or more directional 
events is suggested by the shape 
of the angular distribution of the events. Even combining  
the information on the angular distribution 
with the {\em a priori} expectation for the number of events,  
an interesting hint that Kamiokande~II dataset includes some
elastic scattering events does remain (especially 
if ${\cal E}_x$ is relatively large).

It is conceivable that one can improve the 
agreement between the angular distribution and the 
expected (small) number of elastic scattering events 
by considering non-standard scenarios for the collapse.
In the cases we considered the obtained improvements 
are interesting, but not dramatic.

Let us conclude coming back to the `standard' picture,
and recalling which are the likely values 
of the parameters of the collapse we obtained:
\begin{itemize}
\item In a `standard' picture, we estimated from the data that 
the average energy of $\bar{\nu}_e$ is about 
$E_0=\langle E_{\bar\nu_e}\rangle \sim 14$~MeV. 
This is corroborated in particular 
by the average energy of IMB events and by the fact that
more events are observed at KII. Other pieces of data 
give contrasting hints: the angular distributions
would like to have an average energy as high as possible, 
the average energy at KII suggest instead a lower average energy.
In the `standard' picture, we interpret these features as due 
to fluctuations, possibly with the contribution of 
one or more directional events in the data sets. As for the 
theoretical impact of this result, we note that 
a low average energy suggests an effective thermalization of the emitted 
antineutrinos.
\item The energy emitted in the collapse 
is about ${\cal E}_B\sim 2.5\times 10^{53}$~erg, for a 
distance of 52~kpc. Interestingly, 
this value is not far from simple minded 
theoretical expectations.
Assuming further
long wavelength oscillations in mirror neutrinos 
as in \cite{nb9}
half of the neutrinos become invisible and 
${\cal E}_B$ should be doubled.
Unfortunately, the calculations 
of ${\cal E}_B$ do not seem to be precise enough to
tell $(2-3)$ from $(4-6)\times 10^{53}$~erg at present. 
\item From the hint for elastic scattering event(s) we 
have some preference toward a comparably larger value of 
${\cal E}_x> {\cal E}_{\bar e}$.
This is compatible with current expectations, but it is unclear whether
a large amount of $\nu_{\mu\tau}$ radiation (that does not 
produce `neutrino heating' for the delayed scenario) 
can be easily reconciled with the occurrence of the explosion,
especially if this happens during the accretion phase.
\item Finally, it 
should be noted that there are hints (see~\cite{Udalski} and~\cite{Stanek}) 
from astronomy that the 
distance of the Large Magellanic Cloud traditionally used is overestimated.
If the new value of $D=40$~kpc is adopted, the energy 
emitted in neutrinos that we estimated has to be reduced 
by a factor of $(40/52)^2$, namely ${\cal E}\sim 1.5\times 10^{53}$~erg.
Having little energy at our disposal is unlikely to 
help the occurrence of the supernova. This  
leads us to believe that the old determination of the distance 
is the correct one (as a matter of fact, more recent 
works~\cite{feast,distance} argue from astronomical 
considerations that this is the case). 

\end{itemize}

To summarize our findings, we conclude that the 
`standard' picture of neutrino emission and oscillations is not 
contradicted by SN1987A, and even more, the observed 
properties of the collapse seem to meet expectations. 
We believe that there is wide space for deviations 
from this picture, not only in consideration of the limited statistics
but also due to certain features of the observed signals.
From the discussion (and also in view of other considerations
 \cite{burrows_collapse}), it is evident that there is a great interest in 
obtaining larger samples of elastic scattering events 
and also of events due to $\nu_e$ from the next galactic supernova.

\begin{acknowledgments}
We thank for precious hints, useful discussions  and plain help
V.S.~Berezinsky, F.~Cavanna, M.~Cirelli, V.S.~Imshennik,
W.~Fulgione, E.~Lisi, C.~Pe\~{n}a-Garay,
O.G.~Ryazhskaya and A.~Strumia.
\end{acknowledgments}


\begin{thebibliography}{3}
\bibitem{nb5}
In~\cite{Olga} 
it was observed that, since the energy of directional events is 
greater than 20~MeV, the contribution
to ES should come from $\nu_{\mu,\tau} e^-$ interactions. 
As a consequence, the estimated energy radiated in neutrinos 
is too high. Here, we have shown that this problem is to some extent 
solved by neutrino oscillations.
\end{thebibliography}
\end{document}